\documentclass{llncs}

\usepackage{epsfig}

\title{Decision tree modeling with relational views}

\author{Fadila Bentayeb \and J\'er\^ome Darmont}

\institute{ERIC -- Universit\'e Lumi\`ere Lyon 2\\
5 avenue Pierre Mend\`es-France\\
69676 Bron Cedex \\ France\\
{\email{\{bentayeb | jdarmont\}@eric.univ-lyon2.fr}
}}

\date{}

\begin{document}

\maketitle

\begin{abstract}

Data mining is a useful decision support technique that can be
used to discover production rules in warehouses or corporate data.
Data mining research has made much effort to apply various mining algorithms
efficiently on large databases. However, a serious
problem in their practical application is the long processing time of such algorithms.
Nowadays, one of the key
challenges is to integrate data mining methods within the
framework of traditional database systems. Indeed, such implementations
can take advantage of the efficiency provided by SQL
engines.

In this paper, we propose an integrating approach for decision
trees within a classical database system. In other
words, we try to discover knowledge from relational databases,
in the form of production rules,
via a procedure embedding SQL queries.
The obtained decision tree is
defined by successive, related relational views. Each view corresponds to a
given population in the underlying decision tree. We selected the
classical Induction Decision Tree (ID3) algorithm to build the decision tree.
To prove that our implementation of ID3
works properly, we successfully compared
the output of our procedure with the output of an existing and validated data
mining software, SIPINA.
Furthermore, since our approach is tuneable, it can be generalized to any other
similar decision tree-based method.

\vspace*{0.2cm} {\bf Keywords:}  Integration, Databases, Data Mining,
Decision trees, Relational views.
\end{abstract}

\section{Introduction}

Recently, an important research effort has been made to
apply data mining operations efficiently on large
databases. Indeed, data mining tool vendors  tend to
integrate more and more database features in their products.
However, in practice, the long
processing time required by data mining algorithms remains a critical issue.
Current systems consume minutes or even hours to answer simple mining queries
on very large databases.
On the other hand, database vendors recently began to integrate data mining methods
in the heart of their systems.
Hence, integrating data mining
algorithms within the framework of traditional database systems \cite{Chaudhuri98} becomes
one of the key challenges for research in both the databases and data mining fields.

A first step in this integration process has been achieved by the
rise of data warehousing, whose primary purpose is decision
support rather than reliable storage. A closely related area is
called On-Line Analytical Processing (OLAP) \cite{Codd93}. There
has also been an impressive amount of work related to association
rules, their generalization, and their scalability \cite{MPC96,STA98}.
Relatively, less work has been done in the context
of other classical data analysis techniques from the machine
learning field, e.g., clustering or classification. In this area, most research 
focused on scaling data mining techniques to work with large data sets \cite{Agrawal96,GRV98}.

To truly integrate data mining methods into their systems, database vendors
recently developed extensions to SQL and Application Programming Interfaces (APIs)
\cite{IBM01,OLEDB00,Oracle01,sqlserver2000}.
These tools allow client applications to explore and
manipulate existing mining models and their applications
through an interface similar to that used for exploring tables,
views and other first-class relational objects.

In this paper, we propose to integrate classical data analysis techniques
(namely, decision tree-based methods) within relational database systems.
To achieve this goal, we only use existing structures, namely,
relational views that we exploit through SQL queries. 

To illustrate our approach, we chose to integrate the ID3 
decision tree-based method \cite{QUI86}, which is a supervised
learning method generating knowledge in a production rule-set form.
We selected ID3 mainly because it is quite simple to
implement. However, we plan to
take other, more elaborate methods into account, since our
approach is now validated.

Such algorithms as ID3 generate a decision tree that is a succession of
smaller and smaller partitions of an initial training set. Our key
idea comes from this very definition. Indeed, we can make an
analogy between building successive, related partitions (different
populations) and creating successive, related relational views. Each node of the
decision tree is associated with the corresponding view.
Since SQL database management systems provide a rich set of
primitives for data retrieval, we show  that data mining
algorithms can exploit them efficiently, instead of developing all requirement
functionality from scratch.

To achieve this goal, we designed a PL/SQL stored procedure
that uses SQL queries to generate the decision tree.
Note that the views that are successively created can be stored and thus queried or
analyzed after the tree is generated, if needed.
The main differences between our approach and the existing ones
are:
(1) existing methods extend SQL to support mining operators
  when our approach only uses existing SQL statements and structures;
(2) existing methods use APIs when our approach does not;
and (3) existing methods store the obtained mining models into an extended relational table as in \cite{NCBF00}.
In our approach, the mining model we obtain is defined as a traditional
table representing the decision tree and a set of successive, related views
representing the nodes of the tree.  

The remainder of this paper is organized as follows. 
Section~\ref{section:
our_approach} explains the principle of our approach.
Section~\ref{implementation}
details our implementation of ID3 and the functionality of our stored procedure.
Section~\ref{section:
tests and results} presents the experiments we performed to validate our
approach. We finally conclude this paper and discuss research perspectives in Section~\ref{section:
Perspectives}.

\section{Principle of our approach}
\label{section: our_approach}

Induction
graphs are data mining tools that produce "if-then"-like rules.
They take as input a set of objects (tuples, in the
relational databases vocabulary)
described by a collection of
attributes. Each object belongs to one
of a set of mutually exclusive classes. The induction task determines
the class of any object from the values of its attributes. A training
set of objects whose class is known is needed to build the induction
graph. Hence, an induction graph building method takes as input a set
of objects defined by predictive attributes and a class attribute,
which is the attribute to predict.

\begin{figure}[hbt]
\begin{center}
\epsfxsize=10cm \centerline{\epsffile{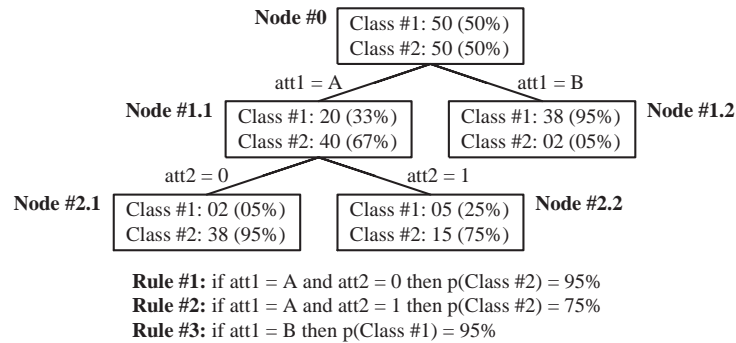}}
\caption{Example of decision tree}
\label{treex}
\end{center}
\end{figure}

Then, these methods apply successive criteria on the training
population to obtain groups wherein the size of one class is
maximized. This process builds a tree, or more generally a graph.
Rules are then produced by following the paths from
the root of the tree (whole population) to the different leaves
(groups wherein the one class represents the majority in the population
strength).
Figure~\ref{treex} provides an example of decision tree with its associated
rules, where p(Class \#i) is the probability of objects to belong to Class \#i.

In our approach, the root node of the decision tree is represented by a relational view
corresponding to
the whole training dataset. Since each sub-node in the decision tree
represents a sub-population of its parent node, we build for each node
a relational view that is based on its parent view. Then, these views
are used to count the population strength of each class in the node
with simple {\small {\tt GROUP BY}} queries.
These counts are used to determine the
criteria that helps either partitioning the current node into a set of disjoint
sub-partitions based on the values of a specific attribute or concluding
that the node is a leaf, i.e., a terminal node.
To illustrate how these views are created, we represented in Figure~\ref{treexviews}
the SQL statements for creating the views associated with the sample decision tree
from Figure~\ref{treex}. This set of views constitutes the decision tree. 

\begin{figure}[hbt]
\begin{center}
\begin{tabular}{|p{11.8cm}|}
\hline {\tt
Node \#0:~~ CREATE VIEW v0 AS SELECT att1, att2, class FROM training\_set \newline
Node \#1.1: CREATE VIEW v11 AS SELECT att2, class FROM v0 WHERE att1='A' \newline
Node \#1.2: CREATE VIEW v12 AS SELECT att2, class FROM v0 WHERE att1='B' \newline
Node \#2.1: CREATE VIEW v21 AS SELECT class FROM v11 WHERE att2=0 \newline
Node \#2.2: CREATE VIEW v22 AS SELECT class FROM v11 WHERE att2=1
}\\
\hline
\end{tabular}
\caption{Relational views associated with sample decision tree}
\label{treexviews}
\end{center}
\end{figure}

\section{Implementation}
\label{implementation}

We used Oracle 8i to implement the ID3 method, under the form of a
PL/SQL stored procedure named {\tt BuildTree}. Its full, commented
code, installation and de-installation scripts, the sample
datasets we used to validate our code, and a short user manual are
freely available on-line\footnote{{\tt
http://eric.univ-lyon2.fr/$\sim$jdarmont/download/buildtree.zip}}.

\subsection{Data structures}

To build a decision tree, we need to manage: (1) the nodes of the tree;
and (2) the candidate attributes for splitting a node, and the associated new nodes each attribute would generate.
Hence, we implemented the following data structures.

\subsubsection{Stack of nodes.}

The stack structure for nodes is justified by the fact we
encountered system errors when coding our tree building procedure
recursively. Hence, we handled recursivity ourselves with a stack.

An individual node is a record composed of the following fields:
{\tt num}, node number;
{\tt nview}, name of the relational view that is associated with the node;
{\tt rule}, the explicit rule that lead to the creation of the node, e.g., {\small {\tt GENDER=FEMALE}} (this is only stored for result output);
{\tt entrop}, node entropy (which variation expresses the discriminating power of an attribute); and
{\tt pop}, node population strength.

\subsubsection{List of candidates.}

Our list of candidates must contain a set of
attributes, the information gain associated with these attributes (expressed as a difference in entropy weighted averages),
and a list of the nodes that would be generated if the current
attribute was selected for splitting the current node. Hence, we
need a list of lists. Such a data structure is impossible to
achieve with the usual PL/SQL collections. The solution we adopted
in this first implementation is using the extented relational features
of Oracle. We used a relational table as our
principal list, with an embedded table (collection) as the list of
nodes.

As a consequence, our table of candidates is composed of the
following fields:
{\tt att\_name}, considered attribute name;
{\tt gain}, information gain; and
{\tt nodes}, embedded list of associated nodes.

\subsection{Algorithm}

\subsubsection{Input parameters.}

The input parameters of our algorithm are given in
Table~\ref{tab1}.

\begin{table}[hbtp]
\begin{center}
\begin{tabular}{|l|c|c|}
\hline {\bf Parameter} & {\bf Name} & {\bf Default value} \\
\hline \hline Data source table name & {\tt table\_name} & --- \\
\hline Class attribute (attribute to predict) & {\tt class} & --- \\
\hline Result table
name & {\tt res\_name} & {\tt BTRES} \\ \hline (Strict) minimum
information gain for node building & {\tt min\_gain} & {\tt 0} \\
\hline Root node view name & {\tt root\_view} & {\tt BTROOT} \\
\hline Clean-up views after execution (True/False) & {\tt del} &
{\tt TRUE} \\ \hline
\end{tabular}
\vspace{0.3cm}
\caption{Algorithm input parameters} \label{tab1}
\end{center}
\end{table}

\subsubsection{Pseudo-code.}

We suppose we can call a procedure named {\tt Entropy()} that
computes both the entropy and the population strength of a node.
These data are used when computing the information gain. {\tt Entropy()}
has actually been coded in PL/SQL. Our algorithm pseudo-code for
the {\tt BuildTree} procedure is provided in Figure~\ref{fig1}.

\begin{figure}[hbt]
\begin{center}
\begin{tabular}{|p{11.2cm}|}
\hline
 {\tt
Create result table \newline Create root node using the data
source table \newline Compute root node entropy and population
strength \newline Push root node \newline Update result table with
root node \newline \underline{While} the stack is not empty
\underline{do} \newline
    \hspace*{0.5cm} Pop current node \newline
    \hspace*{0.5cm} Clean candidate list \newline
    \hspace*{0.5cm} \underline{For} each attribute but the class attribute \underline{do} \newline
        \hspace*{1cm} Create a new candidate \newline
        \hspace*{1cm} \underline{For} each possible value of current attribute \underline{do} \newline
            \hspace*{1.5cm} Build new node and associated relational view \newline
            \hspace*{1.5cm} Compute new node entropy and population strength \newline
            \hspace*{1.5cm} Update information gain \newline
            \hspace*{1.5cm} Insert new node into current candidate node list \newline
        \hspace*{1cm} \underline{End for} (each value) \newline
    \hspace*{0.5cm} \underline{End for} (each attribute) \newline
    \hspace*{0.5cm} Search for maximum information gain in candidate list \newline
    \hspace*{0.5cm} \underline{For} each candidate \underline{do} \newline
        \hspace*{1cm} \underline{If} current attribute bears the greater information gain \underline{then} \newline
            \hspace*{1.5cm} \underline{For} each node in the list of nodes \underline{do} \newline
                \hspace*{2cm} Push current node \newline
                \hspace*{2cm} Update result table with current node \newline
            \hspace*{1.5cm} \underline{End for} (each node) \newline
        \hspace*{1cm} \underline{Else} \newline
            \hspace*{1.5cm} \underline{For} each node in the list of nodes \underline{do} \newline
                \hspace*{2cm} Destroy current node \newline
            \hspace*{1.5cm} \underline{End for} (each node) \newline
        \hspace*{1cm} \underline{End if} \newline
    \hspace*{0.5cm} \underline{End for} (each candidate) \newline
\underline{End while} (stack not empty)
}\\
\hline
\end{tabular}
\caption{Pseudo-code for the {\tt BuildTree} stored procedure}
\label{fig1}
\end{center}
\end{figure}

\subsection{Result output}

The output of our stored procedure, namely a decision tree, is
stored into a relational table whose name is specified as an input
parameter. The table structure reflects the hierarchical structure
of the tree. Its fields are:
{\tt node}, node ID number (primary key, root node is always \#0 --- note that there is a direct link
between the node ID and the associated view name);
{\tt parent}, ID number of parent node in the tree (foreign key, references a node ID number);
{\tt rule}, the rule that lead to the creation of this node, e.g., {\small {\tt GENDER=FEMALE}}; and
for each value V of attribute E, a field labelled {\small {\tt E\_V}}, population strength for the considered value of the attribute in this node.

Such a table is best queried using Oracle SQL hierarchical
statements. The result is directly a textual description of the
output decision tree. A sample query is provided in
Figure~\ref{fig2}.
From this representation, it is very easy to deduce the corresponding
set of production rules.

\begin{figure}[hbt]
\begin{center}
\begin{tabular}{|p{9.55cm}|}
\hline {\tt SELECT LEVEL, node, parent, rule, E\_1, E\_2, ...
FROM btres \newline CONNECT BY node = parent
START WITH node = 0
}\\
\hline
\end{tabular}
\caption{Hierarchical SQL query for decision tree display}
\label{fig2}
\end{center}
\end{figure}

\section{Tests and results}
\label{section: tests and results}

The aim of these experiments is to prove our implementation of the
ID3 decision tree generation method functions properly. For this
sake, we compared the output of our procedure with the output of a
validated data mining tool, SIPINA \cite{ZIG96}, which can be
configured to apply ID3 as well, on several datasets. Due to space
constraints, we only present here our most significant experiment.
However, the full range of our experiments is available 
on-line$^{\mbox{\tiny{1}}}$.

The dataset we selected is
designed to test decision tree building methods. It is aimed
at predicting which classes of passengers of the Titanic are more
likely to survive the wreck. The attributes are:
{\small {\tt CLASS = \{1ST | 2ND | 3RD | CREW\}}};
{\small {\tt AGE = \{ADULT | CHILD\}}};
{\small {\tt GENDER = \{FEMALE | MALE\}}}; and
{\small {\tt SURVIVOR = \{NO | YES\}}} (class attribute).
There are 2201 tuples. Since the {\small {\tt CLASS}} attribute has four
modalities (distinct values), it can generate numerous nodes, and
thus a relatively dense tree.

The results provided by
our procedure, {\tt BuildTree}, are provided in Figure~\ref{fig8}.
Note that we
added in our result query the computation of the relative
populations in each node (in percentage).
Due to the tree width, the results provided by
SIPINA are split-up in Figures \ref{fig9} and \ref{fig10}.
The common point in these two figures is the root node.
As expected, the results provided by SIPINA and {\tt BuildTree}
are the same. 

\begin{figure}[hbt]
\begin{center}
\begin{tabular}{|p{10.8cm}|}
\hline {\tt LEVEL NODE PARENT RULE~~~~~~~~~ SURVIVOR\_NO P\_NO
SURVIVO\_YES P\_YES \newline
----- ---- ------ ------------- ----------- ---- ---------- ----- \newline
1~~~~~~~~0~~~~~~~~~~~~~~~~~~~~~~~~~~~~~1490
~68\%~~~~~~~~711~~~32\% \newline 2~~~~~~~~1~~~~~~0
GENDER=FEMALE~~~~~~~~~126 ~27\%~~~~~~~~344~~~73\% \newline
3~~~~~~~13~~~~~~1 CLASS=CREW~~~~~~~~~~~~~~3
~13\%~~~~~~~~~20~~~87\%          \newline 3~~~~~~~14~~~~~~1
CLASS=1ST~~~~~~~~~~~~~~~4 ~~3\%~~~~~~~~141~~~97\% \newline
4~~~~~~~21~~~~~14 AGE=CHILD~~~~~~~~~~~~~~~0
~~0\%~~~~~~~~~~1~~100\% \newline 4~~~~~~~22~~~~~14
AGE=ADULT~~~~~~~~~~~~~~~4 ~~3\%~~~~~~~~140~~~97\% \newline
3~~~~~~~15~~~~~~1 CLASS=2ND~~~~~~~~~~~~~~13
~12\%~~~~~~~~~93~~~88\% \newline 4~~~~~~~19~~~~~15
AGE=CHILD~~~~~~~~~~~~~~~0 ~~0\%~~~~~~~~~13~~100\% \newline
4~~~~~~~20~~~~~15 AGE=ADULT~~~~~~~~~~~~~~13
~14\%~~~~~~~~~80~~~86\% \newline 3~~~~~~~16~~~~~~1
CLASS=3RD~~~~~~~~~~~~~106 ~54\%~~~~~~~~~90~~~46\% \newline
4~~~~~~~17~~~~~16 AGE=CHILD~~~~~~~~~~~~~~17
~55\%~~~~~~~~~14~~~45\% \newline 4~~~~~~~18~~~~~16
AGE=ADULT~~~~~~~~~~~~~~89 ~54\%~~~~~~~~~76~~~46\% \newline
2~~~~~~~~2~~~~~~0 GENDER=MALE~~~~~~~~~~1364
~79\%~~~~~~~~367~~~21\% \newline 3~~~~~~~~3~~~~~~2
CLASS=CREW~~~~~~~~~~~~670 ~78\%~~~~~~~~192~~~22\%
\newline 3~~~~~~~~4~~~~~~2 CLASS=1ST~~~~~~~~~~~~~118
~66\%~~~~~~~~~62~~~34\% \newline 4~~~~~~~11~~~~~~4
AGE=CHILD~~~~~~~~~~~~~~~0 ~~0\%~~~~~~~~~~5~~100\% \newline
4~~~~~~~12~~~~~~4 AGE=ADULT~~~~~~~~~~~~~118
~67\%~~~~~~~~~57~~~33\% \newline 3~~~~~~~~5~~~~~~2
CLASS=2ND~~~~~~~~~~~~~154 ~86\%~~~~~~~~~25~~~14\% \newline
4~~~~~~~~9~~~~~~5 AGE=CHILD~~~~~~~~~~~~~~~0
~~0\%~~~~~~~~~11~~100\%       \newline 4~~~~~~~10~~~~~~5
AGE=ADULT~~~~~~~~~~~~~154 ~92\%~~~~~~~~~14~~~~8\%
\newline 3~~~~~~~~6~~~~~~2 CLASS=3RD~~~~~~~~~~~~~422
~83\%~~~~~~~~~88~~~17\% \newline 4~~~~~~~~7~~~~~~6
AGE=CHILD~~~~~~~~~~~~~~35 ~73\%~~~~~~~~~13~~~27\% \newline
4~~~~~~~~8~~~~~~6 AGE=ADULT~~~~~~~~~~~~~387
~84\%~~~~~~~~~75~~~16\%
}\\
\hline
\end{tabular}
\caption{{\tt BuildTree} result for TITANIC} \label{fig8}
\end{center}
\end{figure}

\begin{figure}[hbt]
\begin{center}
\epsfxsize=12cm \centerline{\epsffile{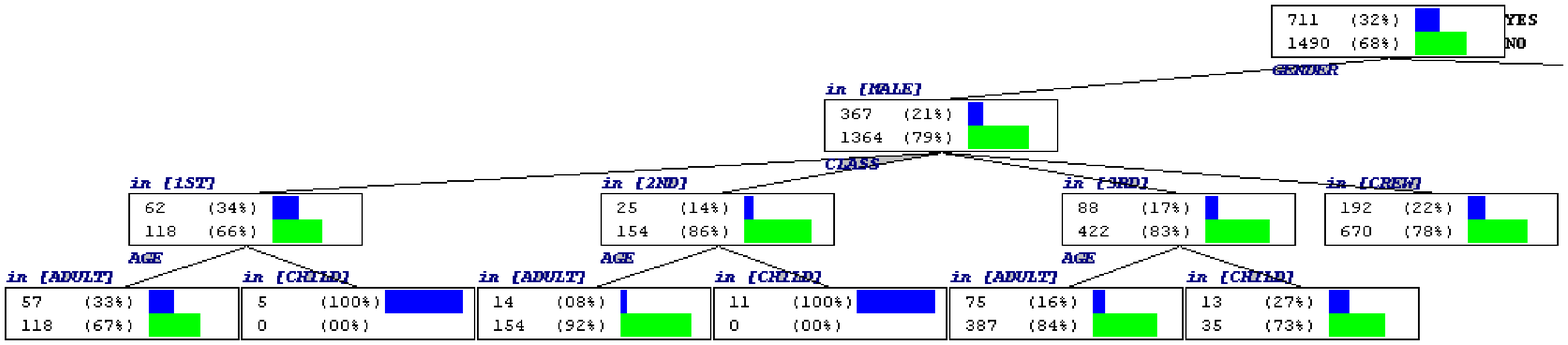}}
\caption{SIPINA result for TITANIC ({\small {\tt GENDER=MALE}})}
\label{fig9}
\end{center}
\end{figure}

\begin{figure}[hbt]
\begin{center}
\epsfxsize=12cm \centerline{\epsffile{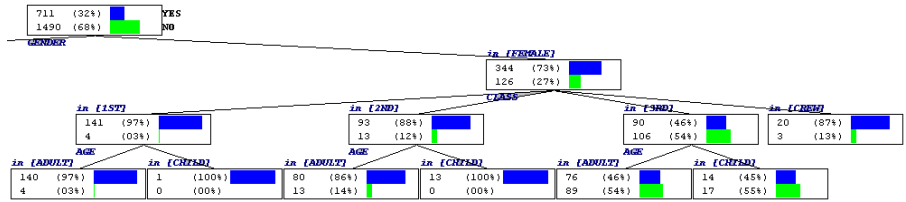}}
\caption{SIPINA result for TITANIC ({\small {\tt GENDER=FEMALE}})}
\label{fig10}
\end{center}
\end{figure}

\section{Conclusion and perspectives}
\label{section: Perspectives}

Major database vendors have all started to integrate data mining features into their systems,
through extensions of the SQL language and APIs.
In this paper, we presented a slightly different approach for integrating data mining operators into
a database system. Namely, we implemented the ID3 method, which we selected for its simplicity,
as a stored procedure that builds a decision tree by
associating each node of the tree with a relational view. 
It is very easy to deduce a set of production rules from the output of our procedure.
This approach has three advantages
over the "black box" tools currently proposed by database vendors:
(1) no extension of the SQL language is needed;
(2) no programming through an API is required;
and (3) the views associated with the nodes of a decision tree can be stored for further analysis
(descriptive statistics or clustering on the sub-population, deployment of a new decision
tree from this node, etc.). The concurrent splitting alternatives could even be retained if
needed.

We sucessfully checked that the results provided by our implementation of ID3 were correct by comparing the output of our
procedure to the output of the SIPINA software, which is a well-known and reliable data mining
platform, on several test datasets of growing complexity.

The perspectives opened by this study are numerous. From a technical point of view,
the performance of our solution could be improved at least at two levels. First, there
is room for code optimization, e.g., by replacing the relational table with an embedded collection
by more efficient, in-memory data structures. Second, a more global optimization scheme could be
achieved by indexing the source table so that building and exploiting the views is faster.

We also need to test the results obtained by {\tt BuildTree} on very large databases.
This would help us determining how well our procedure scales up. We also plan to
compare the performances (i.e., response time) of {\tt BuildTree} and SIPINA on such very large
databases (that do not fit into a computer's main memory) in order to check out that our approach
indeed takes advantage of the host DBMS capabilities.

Eventually, we chose to first implement a very simple decision tree building method (ID3). It
would be interesting to enrich our stored procedure with other, more elaborate methods. Our idea
is to make them available through simple parameterization and keep the tree building as
transparent to the user as possible. We could also integrate other procedures for helping
users to complete the machine learning process, e.g., scoring and cross-validation procedures.

\bibliographystyle{abbrv}

\bibliography{ismis02}

\end{document}